\documentclass[a4paper,12pt]{article}

\titlepage
\usepackage{amssymb}

\usepackage{amsmath}
\usepackage[english]{babel}
\usepackage{url}
   \usepackage[authoryear]{natbib}

\usepackage[utf8]{inputenc}
\usepackage[T1]{fontenc}

\usepackage[pdftex]{graphicx}
\DeclareGraphicsExtensions{.png, .pdf, .jpg} 

\usepackage[pdftex, colorlinks, linkcolor=blue, urlcolor=blue, citecolor=blue, pagecolor=blue, breaklinks=true]{hyperref}

\newcommand{\ds}{\displaystyle}
\newcommand{\ww}{\omega}

\newcommand{\bea}{\begin{eqnarray*}}
\newcommand{\eea}{\end{eqnarray*} }
\newcommand{\be}{\begin{eqnarray}}
\newcommand{\ee}{\end{eqnarray} }
\newtheorem{theorem}{Theorem}[section]
\newtheorem{lemma}{Lemma}[section]
\newtheorem{proposition}{Proposition}[section]
\newtheorem{corollary}{Corollary}[section]
\newtheorem{definition}{Definition}[section]
\newtheorem{example}{Example}[section]
\newtheorem{remark}{Remark}[section]
\newtheorem{statement}{Statement}[section]
\newcommand{\bt}{\begin{theorem}}
\newcommand{\et}{\end{theorem}}
\newcommand{\bl}{\begin{lemma}}
\newcommand{\el}{\end{lemma}}
\newcommand{\bp}{\begin{proposition}}
\newcommand{\ep}{\end{proposition}}
\newcommand{\bc}{\begin{corollary}}
\newcommand{\ec}{\end{corollary}}
\newcommand{\btb}{\begin{table}[hbp]}
\newcommand{\btbh}{\begin{table}[hhh]}
\newcommand{\etb}{\end{table}}
\newcommand{\bfg}{\begin{figure}[hbp]}
\newcommand{\bfgh}{\begin{figure}[hhh]}
\newcommand{\efg}{\end{figure}}
\newcommand{\bd}{\begin{definition}\rm}
\newcommand{\ed}{\end{definition}}
\newcommand{\bex}{\begin{example}\rm}
\newcommand{\eex}{\end{example}}
\newcommand{\br}{\begin{remark}\rm}
\newcommand{\er}{\end{remark}}
\newcommand{\bs}{\begin{statement}}
\newcommand{\es}{\end{statement}}

\begin{document}

\title{ $T$-optimal discriminating  designs  for Fourier \\ regression models}

\author{\small Holger Dette \\
\small Ruhr-Universit\"at Bochum \\
\small Fakult\"at f\"ur Mathematik \\
\small 44780 Bochum, Germany \\
\small e-mail: holger.dette@rub.de\\
\and
\small Viatcheslav B. Melas \\
\small St. Petersburg State University \\
\small Department of Mathematics \\
\small St. Petersburg ,  Russia \\
{\small email: vbmelas@post.ru}\\
\and
\small Petr Shpilev\\
\small St. Petersburg State University \\
\small Department of Mathematics \\
\small St. Petersburg , Russia \\
{\small email: pitshp@hotmail.com}\\
}
\normalsize

\maketitle

\begin{abstract}
In this paper we consider the problem of constructing    $T$-optimal discriminating designs for Fourier regression models. We provide explicit solutions of the optimal design problem for discriminating between two Fourier regression models, which differ by at most three trigonometric functions.
In general, the  $T$-optimal discriminating design depends in a complicated way on the parameters of the larger model, and for special configurations of the parameters
$T$-optimal discriminating designs can  be found analytically. Moreover,   we also study this dependence in the remaining cases by calculating the optimal designs numerically. In particular, it is demonstrated that $D$- and $D_s$-optimal designs have rather low efficiencies with respect to the $T$-optimality criterion.
\end{abstract}

\parindent 0cm
Keywords and Phrases: $T$-optimal design; model discrimination;  linear optimality criteria; Chebyshev polynomial, trigonometric models

AMS subject classification: 62K05

\section{\bf Introduction}
\def\theequation{1.\arabic{equation}}
\setcounter{equation}{0}
 The problem of identifying an appropriate regression  model in a class of competing candidate models is one of the most important problems in applied regression analysis. Nowadays it is well known  that a well designed experiment  can improve the performance of model discrimination substantially, and several authors have addressed the problem of constructing optimal designs for this purpose. The literature on designs for model discrimination 
 can roughly be divided into two parts.
 \cite{hunrei1965}, \cite{stigler1971}  considered two nested models, where
 the extended model reduces to the ``smaller'' model for a specific choice of a subset of the
 parameters. The  optimal discriminating designs are  then  constructed  such that
 these parameters are estimated most precisely. Since these fundamental papers
 several authors have investigated this approach in various regression models [see
  \cite{hill1978}, \cite{studden1982}, \cite{spruill1990}, \cite{dette1994a,dette1995b}, \cite{dethal1998},
 \cite{songwong1999},  \cite{zentsa2004}, \cite{biedethof2009} among many others].
  The second line of research was initialized in a fundamental paper of  \cite{atkfed1975a}, who  introduced the $T$-optimality criterion for discriminating between two competing regression models.  Since the introduction of this criterion, the problem of determining $T$-optimal discriminating designs has been considered by numerous authors [see \cite{atkfed1975b}, \cite{ucibog2005}, \cite{dettit2009}, \cite{atkinson2010}, \cite{tomlop2010} or \cite{wiens2009,wiens2010}  among  others].  The $T$-optimal design problem is essentially a minimax problem, and -- except for very simple models -- the corresponding optimal designs are not easy to find and have to be determined numerically in most cases of practical interest. On the other hand, analytical solutions are helpful for a better understanding of the optimization problem and can also be used to validate numerical procedures for the construction of optimal designs.   Some explicit solutions of the $T$-optimal design problem for discriminating   between two polynomial regression models  can be found in \cite{detmelshp2012}, but to our best knowledge no other analytical solutions are available in the literature.

 In the present paper we consider the problem of constructing $T$-optimal discriminating designs  for   Fourier regression models, which are widely used to describe periodic phenomena [see for example \cite{lestrel1997}].
 Optimal designs for estimating all parameters of the Fourier regression model have been discussed by numerous authors [see
   e.g.
   \cite{karstu1966}, page 347,   \cite{laustu1985},  \cite{kittittor1988}, \cite{ricschwyn1997} and
    \cite{detmel2003}
    among others].
Discriminating design problems   in the spirit of  \cite{hunrei1965}, \cite{stigler1971} have been discussed  by   \cite{biedethof2009}, \cite{zentsa2004} among others, but $T$-optimal designs for   Fourier regression models, have not been investigated in the literature so far. In Section \ref{section2} we introduce the problem and provide a characterization of $T$-optimal discriminating designs in terms of a classical approximation problem. Explicit solutions of the $T$-optimal design problem for Fourier regression models are discussed in Section \ref{section3}. Finally, in Section \ref{section4} we provide some numerical results of these challenging optimization problems. In particular, we demonstrate that the structure (more precisely the number of support points) of the   $T$-optimal discriminating design depends sensitively on the location of the parameters.

\section{$T$-optimal discriminating designs}\label{section2}
\def\theequation{2.\arabic{equation}}
\setcounter{equation}{0}

 Consider the classical regression model
\begin{equation}\label{(1.1)}
   y=\eta(x)+\varepsilon ,
\end{equation}
where the explanatory variable $x$ varies in a compact design space, say $\cal X$, and observations at different locations, say $x$ and $x'$, are assumed to be  independent. In \eqref{(1.1)} the quantity $\varepsilon$ denotes a random variable with mean 0 and variance $\sigma^{2}$, and $\eta(x)$ is a function, which is called regression function in the literature [see \cite{sebwil1989}].
We assume that the experimenter has two parametric models, say $\eta_1(x,\theta_1)$  and $\eta_2(x,\theta_2)$, for this function in mind to describe the relation between predictor and response,
 and that the first goal of the experiment is to identify the appropriate model from these two candidates.
 In order to find ``good'' designs for discriminating between the models $\eta_1$ and $\eta_2$ we consider approximate designs in the
sense of \cite{Kiefer1974},
which are  probability measures on the design space $\mathcal{X}$ with finite support.
The support points, say $x_1,\dots, x_s$, of an (approximate) design $\xi$ define the locations where observations
are taken, while the weights denote the corresponding relative
proportions of total observations to be taken at these points.
If the design $\xi$ has masses $\omega_i>0 $ at the different points $x_i \: (i =
1, \dots, s)$ and $n$ observations can be made,  the quantities
$\omega_i   n$ are rounded to integers, say $n_i$, satisfying $\sum^s_{i=1} n_i =n$, and
the experimenter takes $n_i$ observations at each location $x_i \:  (i=1, \dots, s)$ [see for example
\cite{pukrie1992}]. \\
For the construction of a good design for discriminating between the models $\eta_1$ and $\eta_2$  \cite{atkfed1975a} proposed in a seminal paper to fix
one model, say $\eta_2$, and to determine the discriminating design such that the minimal deviation between the model $\eta_2$ and the class of models defined by $\eta_1$ is maximized. More precisely, a
 $T$-optimal design is defined $\xi^*$  by
\begin{eqnarray*}
\xi^*=\arg \max_{\xi}\int_{\chi}\left(\eta_2(x,\theta_2)-\eta_1
(x,\widehat{\theta}_1)\right)^{2}\xi(d x),
\end{eqnarray*}
where the parameter $\widehat{\theta}_1$ minimizes the expression
\begin{eqnarray*}
\widehat{\theta}_1=\arg \min_{\theta_1}\int_{\chi}
\left(\eta_2(x,\theta_2)-\eta_1(x,\theta_1)\right)^{2}\xi(d x).
\end{eqnarray*}
Note that the $T$-optimality criterion is a local optimality criterion in the sense of \cite{chernoff1953}, because it requires knowledge of the parameter $\theta_2$. Bayesian versions of this criterion have recently been investigated by \cite{detmelshp2013} and \cite{detmelguc2015}.

In the present work we consider cases, where the competing   regression functions are given by two Fourier regression models of different order, that is
\begin{equation}
\label{1.3}
\eta_{1}(x,\theta_{1}) = \overline{q}_{0}+\sum_{i=1}^{k_1}\overline{q}_{2i-1}\sin(ix)+\sum_{i=1}^{k_2}\overline{q}_{2i}\cos(ix)
\end{equation}
and
\begin{eqnarray}
\label{1.4}
 \eta_{2}(x,\theta_{2})&=&\tilde{q}_{0}+\sum_{i=1}^{k_1}\tilde{q}_{2i-1}\sin(ix)+\sum_{i=1}^{k_2}\tilde{q}_{2i}\cos(ix)\\
&+& \sum_{i=k_1+1}^{m}b_{2(i-k_1)-1}\sin(ix)+\sum_{i=k_2+1}^{m}b_{2(i-k_2)}\cos(ix), \nonumber
\end{eqnarray}
where
\begin{eqnarray*}
 \theta_1 &=& (\overline{q}_0, \overline{q}_2, \ldots, \overline{q}_{2k_2}, \overline{q}_1, \ldots, \overline{q}_{2k_1-1}) \\
 \theta_2 &=& (\tilde {q}_0, \ldots, \tilde {q}_{2k_2},  \tilde {q}_1, \ldots, \tilde {q}_{2k-1},
 b_2, \ldots, b_{2m}, b_1, \ldots, b_{2m-1})
\end{eqnarray*}
are the parameter vectors in model $\eta_1$ and $\eta_2$, respectively.
Fourier regression models
are widely used to describe periodic phenomena [see e.g. \cite{mardia1972}, or  \cite{lestrel1997}] and the problem of designing experiments
for Fourier regression models has been discussed by several authors [see the cited references    in the introduction]. However, the problem of constructing
 $T$-optimal discriminating designs for these models has not been addressed in the literature so far.

We assume that the design space is given by the interval $\chi=[0,2\pi]$ and denote the difference
 $\eta_{2}(x,\theta_{2})-\eta_{1}(x,\theta_{1})$ by
\begin{eqnarray}
\label{1.5}
\overline{\eta}(x,q,\overline{b}) &=& q_{0}+\sum_{i=1}^{k_1}q_{2i-1}\sin(ix)+\sum_{i=1}^{k_2}q_{2i}\cos(ix)\\
&& +\sum_{i=k_1+1}^{m}b_{2(i-k_1)-1}\sin(ix)+\sum_{i=k_2+1}^{m}b_{2(i-k_2)}\cos(ix),
\nonumber
\end{eqnarray}
 where $q=(q_0, q_1, \ldots, q_{2k_1-1}, q_2, \ldots, q_{2k_2})$, $q_i=\tilde{q}_{i}-\overline{q}_i$ and $\overline{b}=(b_1,b_3,\ldots,b_{2(m-k_1)-1}, b_2, b_4, \ldots,b_{2(m-k_2)} )^T$ denotes the vector of ``additional'' parameters in   model \eqref{1.4}.  With these notations the  $T$-optimality criterion reduces to
$$
T(\xi,\overline{b})=\min_{q}\int^{2 \pi}_{0}\overline{\eta}^{2}(x,q,\overline{b})\xi(d x),
$$
and a   $T$-optimal  design for discriminating between the models \eqref{1.3} and \eqref{1.4} maximizes $T(\xi,\overline b)$, that is
$$
            \xi^*=\arg \max_{\xi} T(\xi,\overline{b}).
$$
The following  result provides a characterization of  $T$-optimal designs and  is known in the literature as the equivalence theorem for $T$-optimality  [see, for instance, Theorem 2.2 in \cite{dettit2009}].

\bt\label{Theorem3.1}
    For a  fixed vector $\overline{b}$ the following conditions are equivalent:
\begin{itemize}
\item [(1)] The design
\begin{eqnarray*}
    \xi^*=\left(\begin{array}{ccc} x^*_1&\ldots&x^*_n\\
    \omega_1&\ldots&\omega_n\end{array}\right),\,\, x_i\in[0,2\pi],\,\,
    i=1,\ldots,n.
\end{eqnarray*}
is  a   $T$-optimal for discriminating designs for the   models
$\eta_1$ and $\eta_2$.\
\item [(2)] There exists a vector $\theta^{*}$ and a positive constant $h$ such, that the function
$\psi^*(x)=\overline{\eta}(x,\theta^{*},\overline{b})$ satisfies  the following conditions
\begin{itemize}
\item[(i)] $|\psi^*(x)|\leq h$, \quad \mbox{for all} \quad $x\in[0,2\pi]$,
\item[(ii)] $|\psi^*(x_i)|=h$,  \quad \mbox{for all} \quad $i=1,2,\dots,n$.
\item[(iii)] The support points $x^*_i$ and weights $\omega_i$ of the design $\xi^*$ satisfy the  conditions
\begin{equation} \label{equiv}
\sum^n_{i=1}\psi^*(x^*_i)\frac{\partial \overline{\eta}(x^*_i,\theta,\overline{b})}{{\displaystyle \partial \theta_j}}\omega_i \Big |_{\theta = \theta^*}=0,\quad j=0,\ldots,k_1+k_2.
\end{equation}
\end{itemize}
\end{itemize}

\et

Note that Theorem \ref{Theorem3.1} is not restricted to Fourier regression models but  holds in general for   linear models. A detailed discussion can be found in  \cite{dettit2009}.
As pointed out in the introduction, the explicit determination of $T$-optimal discriminating designs is a very challenging problem. The complexity of the problem depends on the dimension of   the vector $\overline b$. In  the following Sections \ref{section3} and \ref{section4}   we provide explicit and numerical solutions of this  difficult optimal design problem  for Fourier regression models. In particular, we will demonstrate that the structure of the $T$-discriminating design (such as the number of support points) depends on the location of the vector $\overline b$ in the $(2m-k_1-k_2)$-dimensional Euclidean space.

\section{Explicit solutions }\label{section3}
\def\theequation{3.\arabic{equation}}
\setcounter{equation}{0}

In this section we give  some  explicit   $T$-optimal discriminating designs for Fourier regression models.
In particular we consider the problem of constructing   $T$-optimal discriminating designs for the models \eqref{1.3}  and \eqref{1.4}, where
\begin{eqnarray}
\label{(i)} &&  k_1=k_2=m-1, \\
\label{(ii)} &&
k_1=m-1,\ k_2=m-2, \\
\label{(iii)} &&
k_1=m-2,\ k_2=m-1.
\end{eqnarray}

We  give an explicit solution for the   case \eqref{(i)}, while  for the case  \eqref{(ii)}
explicit results are provided in Section \ref{Section3b} for specific values of the parameters $b_\ell$ in model \eqref{1.5}. Corresponding results for the case \eqref{(iii)} are briefly mentioned in Remark \ref{remneu}.
 In general the solution of the   $T$-optimal design problem depends in a complicated way on the parameters $\overline{b}$, and
 we demonstrate numerically in Section \ref{section4}  that the number of support points of the $T$-optimal discriminating design
changes  if the  vector $\overline{b}$ is
  located in different areas of the space $\mathbb{R}^2$.

\subsection{Discriminating designs for $k_1=k_2=m-1$}\label{Section3a}
Throughout this section we assume that $k_1=k_2=m-1$ and rewrite the function in (\ref{1.5}) as
\begin{eqnarray*}
   \overline{\eta}(x,q,\overline{b})= q_{0}+\sum_{i=1}^{m-1}q_{2i-1}\sin(ix)+\sum_{i=1}^{m-1}q_{2i}\cos(ix)+b_{1}\sin(mx)+
   b_{2}\cos(mx).
\end{eqnarray*}
Our first result gives an explicit solution of the $T$-optimal design problem in the case $b_1,b_2\neq 0$.

\begin{theorem}
\label{lem3.1}
 Consider the Fourier regression models \eqref{1.3} and \eqref{1.4} with $k_1=k_2=m-1$. Let $b_1,b_2\neq 0$, then the design
  \begin{eqnarray}
\xi^*=\begin{pmatrix}
\frac{1}{m}\arctan(\frac{1}{b}) & \frac{1}{m}\arctan(\frac{1}{b})+\frac{\pi}{m} & \dots & \frac{1}{m}\arctan(\frac{1}{b})+\frac{(2m-1)\pi}{m} \\  \frac{1}{2m} & \frac{1}{2m} & \dots & \frac{1}{2m}
\end{pmatrix}
\label{(3.5)}
\end{eqnarray}
  is a $T$-optimal discriminating design, where $b=b_2/b_1$.
\end{theorem}
{\bf Proof}.
 We consider
  the function
\begin{eqnarray*}
\psi^*(x)=\overline{\eta}(x,0,\overline{b})=b_{1}\sin(mx)+b_{2}\cos(mx)
\end{eqnarray*}
and prove that this function  and the weights $\omega^*_i = \frac {1}{2m}$ and  support points $x_i^* = \frac {1}{m} \arctan (\frac {1}{b}) + \frac {(i-1)}{m}\pi$ of the design $\xi^*$ defined in (\ref{(3.5)}) satisfy   the conditions   of   Theorem \ref{Theorem3.1}.

Direct calculations show for  the  support points of the design $\xi^*$  the identities
$$
\psi^* (x^*_i)= (-1)^{i-1} \sqrt{b^2_1 + b^2_2}, \qquad i=1,\ldots,2m.
$$
Consequently, the function $\psi^*$ satisfies   conditions  (i)-(ii) for $h=\sqrt{b^2_1 + b^2_2}$, and it remains to show that the equations in \eqref{equiv} hold.
In other words, we have to check that the   equalities
\begin{eqnarray} \label{nec}
\sum^n_{i=1}(-1)^i\sin(j x_i^*)=0, \ \ \sum^n_{i=1}(-1)^i\cos(j x_i^*)=0,\ \ j=0,\ldots, m-1,
\end{eqnarray}
are satisfied.
Observing the identities $\sin(\alpha+\beta)= \sin(\alpha)\cos(\beta)+\cos(\alpha)\sin(\beta)$ and
$\cos(\alpha+\beta)= \cos(\alpha)\cos(\beta)-\sin(\alpha)\sin(\beta)$ we can rewrite  \eqref{nec} as
\begin{eqnarray*}
\sum^{2m}_{i=1}(-1)^i\sin \Big(j\frac{(i-1)\pi}{m}\Big)=0, \ \ \sum^{2m}_{i=1}(-1)^i\cos \Big (j\frac{(i-1)\pi}{m} \Big)=0,\ \  j=0,\ldots, m-1.
\end{eqnarray*}
These equalities are  a consequence of the identity
$$
\sum^{2m-1}_{\ell = 0} e^{ \frac {i \pi \ell j}{m}} (-1)^\ell = 0 \qquad j=1,\ldots,m-1
$$
(here $i=\sqrt{-1}$ and the case $j=0$ has to be considered separately), and  the assertion of Theorem \ref{lem3.1}  now follows from Theorem \ref{Theorem3.1}.
\hfill $\Box$

\begin{corollary}
\label{lem3.2}
 Consider the Fourier regression models \eqref{1.3} and \eqref{1.4} with $k_1=k_2=m-1$.
 If $b_1=0$, then the design
 $$
\xi^*=\begin{pmatrix}
0 & \frac{\pi}{m} & \dots & \frac{(2m-1)\pi}{m} \\  \frac{1}{2m} & \frac{1}{2m} & \dots & \frac{1}{2m}
\end{pmatrix}
$$
  is a $T$-optimal discriminating design. If  $b_2=0$, then the design  $$
\xi^*=\begin{pmatrix}
 \frac{\pi}{2m} & \frac{3\pi}{2m} & \dots & \frac{(4m-1)\pi}{2m} \\  \frac{1}{2m} & \frac{1}{2m} & \dots & \frac{1}{2m}
\end{pmatrix}
$$ is a $T$-optimal discriminating design.
\end{corollary}

\begin{figure}[hhh!]
      \begin{center}
          \includegraphics[width=100mm]{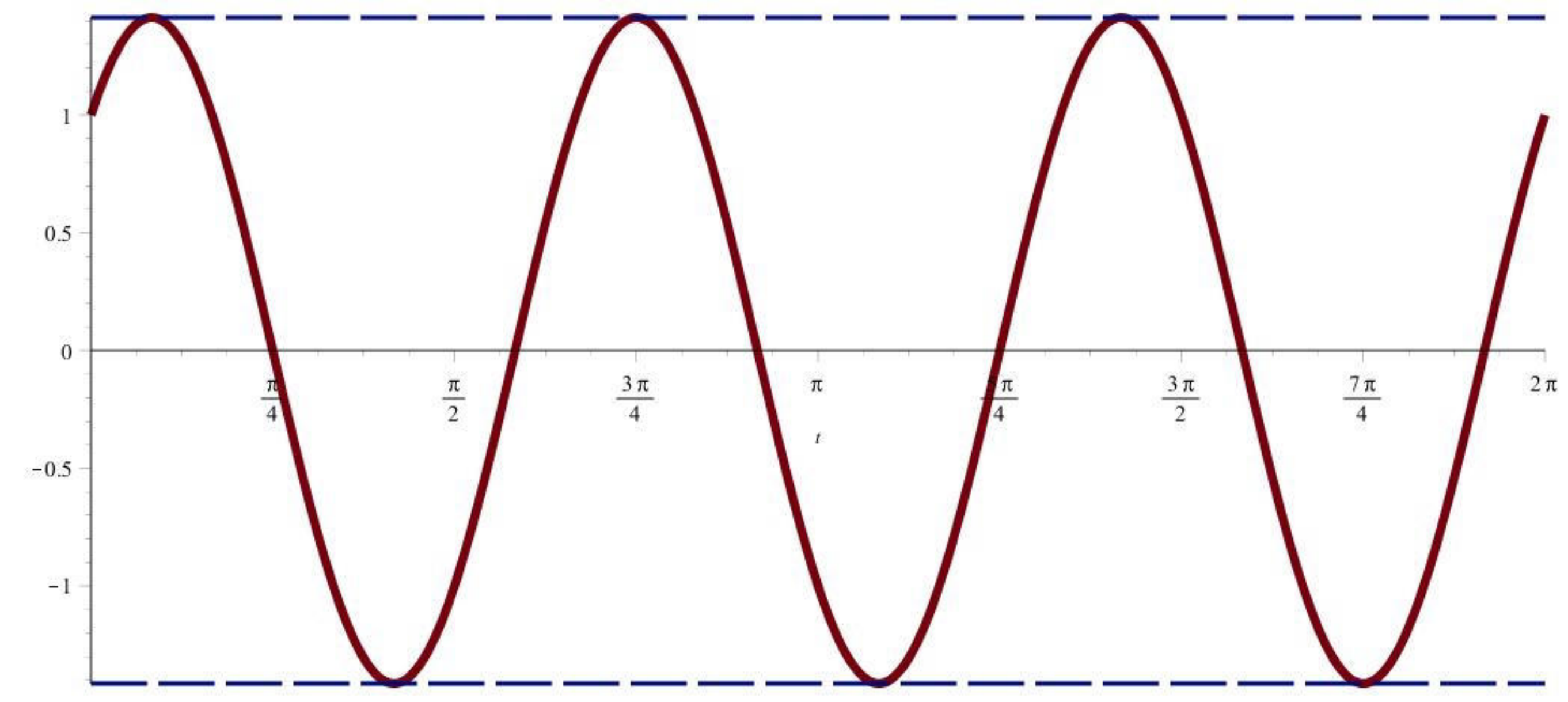}
      \end{center}
       \caption{ \it The   function $\psi^*$ of Theorem \ref{Theorem3.1} for the two Fourier regression models \eqref{ex11} and \eqref{ex12}   ($b_1=b_2=1$).   }
       \label{fig01}
    \end{figure}

\bex
Suppose that $m=3,$ $b_1=b_2=1$ and $k_1$=$k_2$=2, then
if follows from Theorem \ref{lem3.1} that  the design
with equal masses at the six points $\frac{1}{12}\pi$, $ \frac{5}{12}\pi$, $ \frac{3}{4}\pi $, $  \frac{13}{12}\pi $,  $   \frac{17}{12}\pi$
and  $  \frac{7}{4}\pi$ is a  $T$-optimal discriminating design for the two Fourier regression models
\begin{eqnarray}
&& \overline{q}_0 + \overline{q}_1 \sin x + \overline{q}_2 \cos x + \overline{q}_3 \sin (2x)
+ \overline{q}_4 \cos (2x) \label{ex11} \\
&& \overline{q_0} +  \tilde{q}_1 \sin x +  \tilde{q}_2 \cos x +  \tilde{q}_3 \sin (2x)
+  \tilde{q}_4 \cos (2x) + {b_1} \sin (3x) +  {b_2} \cos (3x) . \label{ex12}
\end{eqnarray}
The   function $\psi^*$  of   Theorem \ref{Theorem3.1} for this design is depicted in Figure \ref{fig01}.
 \eex

\subsection{\bf Discriminating designs for $k_1=m-1,k_2=m-2$}\label{Section3b}

Throughout this section we determine $T$-optimal discriminating designs for the trigonometric regression models
\begin{eqnarray} \label{n1}
\eta_1 (x, \theta_1)= &=& \overline{q}_0 + \sum^{m-1}_{i=1} \overline{q}_{2i-1} \sin (ix) + \sum^{m-2}_{i=1} \overline{q}_{2i} \cos (ix) \\
\label{n2}
\eta_2 (x, \theta_2) &=& \tilde{q}_0 + \sum^{m-1}_{i=1} \tilde{q}_{2i-1} \sin (ix) + \sum^{m-2}_{i=1} \tilde{q}_{2i} \cos (ix) \\ \nonumber
&+& b_0 \cos ((m-1)x) + b_1 \sin (mx) + b_2 \cos (mx).
\end{eqnarray}
 Note that the two regression models in \eqref{1.3} and \eqref{1.4} now differ by three functions, that is   $k_1=m-1,k_2=m-2$.
 In general,   $T$-optimal discriminating designs for this case have to be determined numerically, and we will provide  numerical results for  $m=2$ and $m=3$ in the following Section \ref{section4}. However, for some special configurations of the parameters, the  $T$-optimal discriminating designs can also be found explicitly, and these cases will be discussed in the present section.

 If $k_1=m-1,k_2=m-2$  the function $\overline{\eta}$ in (\ref{1.5}) has the representation
 \begin{eqnarray*}
\overline{\eta}(x,q,\overline{b})=  q_{0}+\sum_{i=1}^{m-1}q_{2i-1}\sin(ix)+\sum_{i=1}^{m-2}q_{2i}\cos(ix) +  b_{0}\cos((m-1)x)+b_{1}\sin(mx)+
   b_{2}\cos(mx).
\end{eqnarray*}
We may assume without loss of generality that $b_0 = 1 $.
Indeed, if  $b_0 =0$, the optimal designs can be obtained from Theorem \ref{lem3.1}. Moreover, if $b_0 \neq 0$, the $T$-optimal discriminating design does not depend on the particular value of $b_0$, since we can divide all   coefficients by this parameter.
After normalizing   we therefore obtain
 \begin{eqnarray} \label{2.4a}
\overline{\eta}(x,q,\overline{b}) &=& q_{0}+\sum_{i=1}^{m-1}q_{2i-1}\sin(ix)+\sum_{i=1}^{m-2}q_{2i}\cos(ix) +
  \cos((m-1)x) \\
\nonumber && +b_{1}\sin(mx)+
   b_{2}\cos(mx).
\end{eqnarray}
We now   concentrate   on two special cases: $b_1=0,\ b_2\neq 0$  and $b_2=0,\ b_1\neq 0$,
for which we can provide an explicit solution of the $T$-optimal design problem if the absolute value    of the non-vanishing parameter  is sufficiently large. For this purpose we define support points and weights  as follows
\begin{eqnarray}\label{4.1}
x^*_i (b) =\arccos\Big(-\Big(1+\frac{1}{2m|b|}\Big)\cos\Big(\frac{(m-i+1)\pi}{m}\Big)-\frac{1}{2m|b|}\Big),\
\\
{\ww^*_i=\frac{1}{m}\cos^2\Big(\frac{(i-1)\pi}{2m}\Big), \ i=1,\ldots,m.\hskip9.6em}\label{4.2}
\end{eqnarray}
Our next result gives an explicit solution of the $T$-optimal design problem in the case $b_1=0,\ b_2 \neq 0$.

\bt\label{Theorem4.1}
Consider the Fourier regression models \eqref{n1} and \eqref{n2}  with $b_0=1$,  $b_1=0$, $b_2 \neq 0$.
\begin{itemize}
\item[(a)] If $b_2> 0$,   $|b_2|\geq\frac{1}{2m}\cot^2\left(\frac{\pi}{2m}\right)$, then the design
  \begin{eqnarray}\label{optdes1}
\xi^*_1=\left(\begin{array}{cccccc} x^*_1(b_2)&\ldots & x^*_m (b_2)& 2\pi-x^*_{m}(b_2) &\ldots & 2\pi-x^*_{2}(b_2)\\
    \omega^*_1&\ldots&\omega^*_{m}&\omega^*_{m}& \ldots&\omega^*_2 \end{array}\right)
   \end{eqnarray}
is a  $T$-optimal discriminating   design, where the support points
and weights are defined in (\ref{4.1}) and (\ref{4.2}), respectively.
    \item[(b)] If $b_2<0$,   $|b_2|\geq\frac{1}{2m}\cot^2\left(\frac{\pi}{2m}\right)$, then the design
  \begin{eqnarray} \label{optdes2}
  \xi^*_2=\left(\begin{array}{cccccc} \pi-x^*_m(b_2)&\ldots & \pi-x^*_1(b_2) & \pi+x^*_{2} (b_2) &\ldots & \pi+x^*_{m}(b_2)\\
    \omega^*_m&\ldots&\omega^*_{1}&\omega^*_{2}& \ldots&\omega^*_m \end{array}\right),
\end{eqnarray}
is a   $T$-optimal discriminating   design, where the support points
and weights are defined in (\ref{4.1}) and (\ref{4.2}), respectively.
\end{itemize}
\et

{\bf Proof}. We only consider the case $b_2 \geq\frac{1}{2m}\cot^2(\frac{\pi}{2m})>0$ and note that the
other case follows by similar arguments. We will use  Theorem \ref{Theorem3.1} and prove  the existence of a vector $\theta^{*}$,
such that the function $\psi^*(x)=\overline{\eta}(x,\theta^{*},\overline{b})$ satisfies   conditions (i) - (iii) in this theorem. For this
purpose let  $T_m (x) = \cos (m \arccos x) $ denote the $m$th Chebyshev polynomial of the first kind, then it is follows by a
straightforward calculation that
there exists  a vector $\theta^{*}$ such that the function
\begin{eqnarray*}
\psi^*(x)=\overline{\eta}(x,\theta^{*},\overline{b})=(-1)^{m} |b_2|\Big(1+\frac{1}{2m|b_2|}\Big)^{m}T_m\Big(\frac{-\cos(x)-\frac{1}{2m|b_2|}}{1+\frac{1}{2m|b_2|}}\Big),
\end{eqnarray*}
is a trigonometric polynomial of degree $m$ with leading term $|b_2| \cos (mx)$ [note that the leading term of $T_m(x) $ is given by $ 2^{m-1}x^m$ and that
$2^{m-1} (\cos x)^{m} =  \cos (mx) + m \cos ((m-2)x)+ \ldots $]. Direct calculations show that the points $x^*_i (b_2)$ defined in \eqref{4.1} are the extremal points of this function,
that is
\begin{equation} \label{extremal}
\psi^*(x^*_i(b_2) ) =  (-1)^{i-1} |b_2|\Big(1+\frac{1}{2m|b_2|}\Big)^{m}~, \qquad i=1,\ldots,m.
\end{equation}
Consequently,   $\psi^*$ satisfies  conditions (i) and (ii) of  Theorem \ref{Theorem3.1}.
Finally, we  prove  the third condition \eqref{equiv}. The corresponding equalities reduce to
\begin{eqnarray} \label{eq1}
 {\ds \sum^m_{i=1}\omega^*_i \psi^*(x^*_i (b_2) ) \cos(jx^*_i(b_2))}
 + {\ds \sum^m_{i=2}\omega^*_i \psi^*(2\pi-x^*_i (b_2) ) \cos\big(j(2\pi-x^*_i (b_2))\big)=0}
\end{eqnarray}
$(j=0,\ldots,m-2)$, and
\begin{eqnarray}
\label{eq2}
{\ds \sum^m_{i=1}\omega^*_i \psi^*(x^*_i(b_2)) \sin(jx^*_i (b_2)) }
+  {\ds\sum^m_{i=2}\omega^*_i \psi^*(2\pi-x^*_i(b_2)) \sin\big(j(2\pi-x^*_i (b_2)\big)=0}
\end{eqnarray}
$(j=1,\ldots,m-1)$. Observing \eqref{extremal}, $x^*_1(b_2)=0$ and the identity $\psi^*(x)=\psi^*(2\pi-x)$ we can rewrite
the left hand side of
\eqref{eq2} as
\begin{eqnarray}
&&\sum^m_{i=2}\omega^*_i \big(\sin(jx^*_i (b_2) )+\sin(j(2\pi-x_i^*  (b_2)) )\big)=\sum^m_{i=2} \omega^*_i \big(\sin(jx^*_i (b_2) )-\sin(jx^*_i (b_2) )\big)
\nonumber
\end{eqnarray}
$(j=1,\ldots,m-1)$. Consequently \eqref{eq2}  is obviously satisfied. Similarly, we obtain for \eqref{eq1}
\begin{eqnarray}
&& \sum^m_{i=1}\overline{\omega}_i(-1)^i \cos(jx^*_i (b_2) )=0,\ \ j=0,\ldots,m-2, \label{trig} \
\end{eqnarray}
where  we use the notations $\overline{\omega}_1=\frac{\omega^*_1}{2}$, $\overline{\omega}_i=\omega^*_i,\ i=0,\ldots,m-2$
in \eqref{trig}.
Defining $t_i  = \cos (x^*_i(b_2))$  we obtain for the left hand side of \eqref{trig}
\begin{eqnarray*}
\sum^m_{i=1}\overline{\omega}_i(-1)^i \cos(jx^*_i(b_2) )=\sum^m_{i=1}\sum^{m-2}_{p=0}\overline{\omega}_i(-1)^i a_pt_i^p=\sum^{m-2}_{p=0}a_p\sum^m_{i=1}\overline{\omega}_i(-1)^i t_i^p
\end{eqnarray*}
for some coefficients $a_p$.  It is proved in  Appendix  A.1 of \cite{detmelshp2012} that
\begin{eqnarray*}
\sum^m_{i=1}\overline{\omega}_i(-1)^i t_i^p=0, ~~p=0, \ldots, m-2
\end{eqnarray*}
which implies \eqref{eq1}. The  $T$-optimality of the design $\xi^*_1$ now  directly follows from Theorem \ref{Theorem3.1}. \hfill $\Box$

\bigskip

The next theorem  considers the case $b_1\neq 0,\ b_2 = 0$, which is substantially harder.
Here we are able to determine the   $T$-optimal discriminating designs explicitly if the degree $m$ of the Fourier regression model is odd.

\bt\label{Theorem4.2}
Consider the Fourier regression models \eqref{n1} and \eqref{n2} with $b_0=1, b_1 \neq 0, b_2=0$, where  $m$ is odd. For $\ell =1,2$ let $t^{(\xi_\ell)}_i$   and $\omega^{(\xi_\ell)}_i$,
denote the support points and weights of the designs $\xi^*_1$ and $\xi^*_2$ defined in \eqref{optdes1} and \eqref{optdes2}  
and define
 $$
  t^{(\ell)}_i=t^{(\xi_\ell)}_i+\frac{\pi}{2}\ mod\ 2\pi ; ~~ \ell =1,2.
 $$
 \begin{itemize}
 \item[(a)]
If   $b_1   \geq\frac{1}{2m}\cot^2\left(\frac{\pi}{2m}\right)$, then the design
  \begin{eqnarray}\label{neu1}
 \widetilde{\xi}^*_1=\left(\begin{array}{ccc} t^{(1)}_1&\ldots & t^{(1)}_{2m-1}\\
    \omega^{(\xi_1)}_1&\ldots&\omega^{(\xi_1)}_{2m-1} \end{array}\right)
   \end{eqnarray}
is a   $T$-optimal discriminating design.
 \item[(b)]   If   $b_1<0$,
  $|b_1|\geq\frac{1}{2m}\cot^2\left(\frac{\pi}{2m}\right)$, then the design
  \begin{eqnarray} \label{neu2}
\widetilde{\xi}^*_2=\left(\begin{array}{ccc} t^{(2)}_1&\ldots & t^{(2)}_{2m-1}\\
    \omega^{(\xi_2)}_1&\ldots&\omega^{(\xi_2)}_{2m-1} \end{array}\right)
\end{eqnarray}
is a   $T$-optimal discriminating design.
\end{itemize}
\et

{\bf Proof.} The proof is similar to the proof of Theorem \ref{Theorem4.1},
where we use the   function
\begin{eqnarray*}
\psi^*(x)=\overline{\eta}(x,\theta^{*},\overline{b})=(-1)^{\frac{m+1}{2}} |b_1|\Big(1+\frac{1}{2m|b_1|}\Big)^{m}T_m\Big(\frac{-\sin(x)-\frac{1}{2m|b_1|}}{1+\frac{1}{2m|b_1|}}\Big),
\end{eqnarray*}
in Theorem \ref{Theorem3.1}
The fact that this function is of the form
$$
q_0 + \sum^{2d-2}_{i=1} q_{2i-1} \sin (ix) + \sum^{2d-3}_{i=1} q_{2i} \cos (ix) + \cos ((2d-2)x) + b_1 \sin((2d-1)x)
$$
and satisfies the assumptions of Theorem
\ref{Theorem3.1} follows  from the identity
\begin{eqnarray} \label{arcos}
\cos\big((2d-1)  \arccos(t)\big)\equiv(-1)^{d-1} \sin\big((2d-1)\arcsin(t)\big),\ t\in[-1,1],\ d=1,2, \ldots ,
\end{eqnarray}
which can be used in the case $m=2d-1$. The details are omitted
for the sake of brevity. \hfill $\Box$

\begin{figure}[hhh!]
      \begin{center}
       \includegraphics[width=70mm]{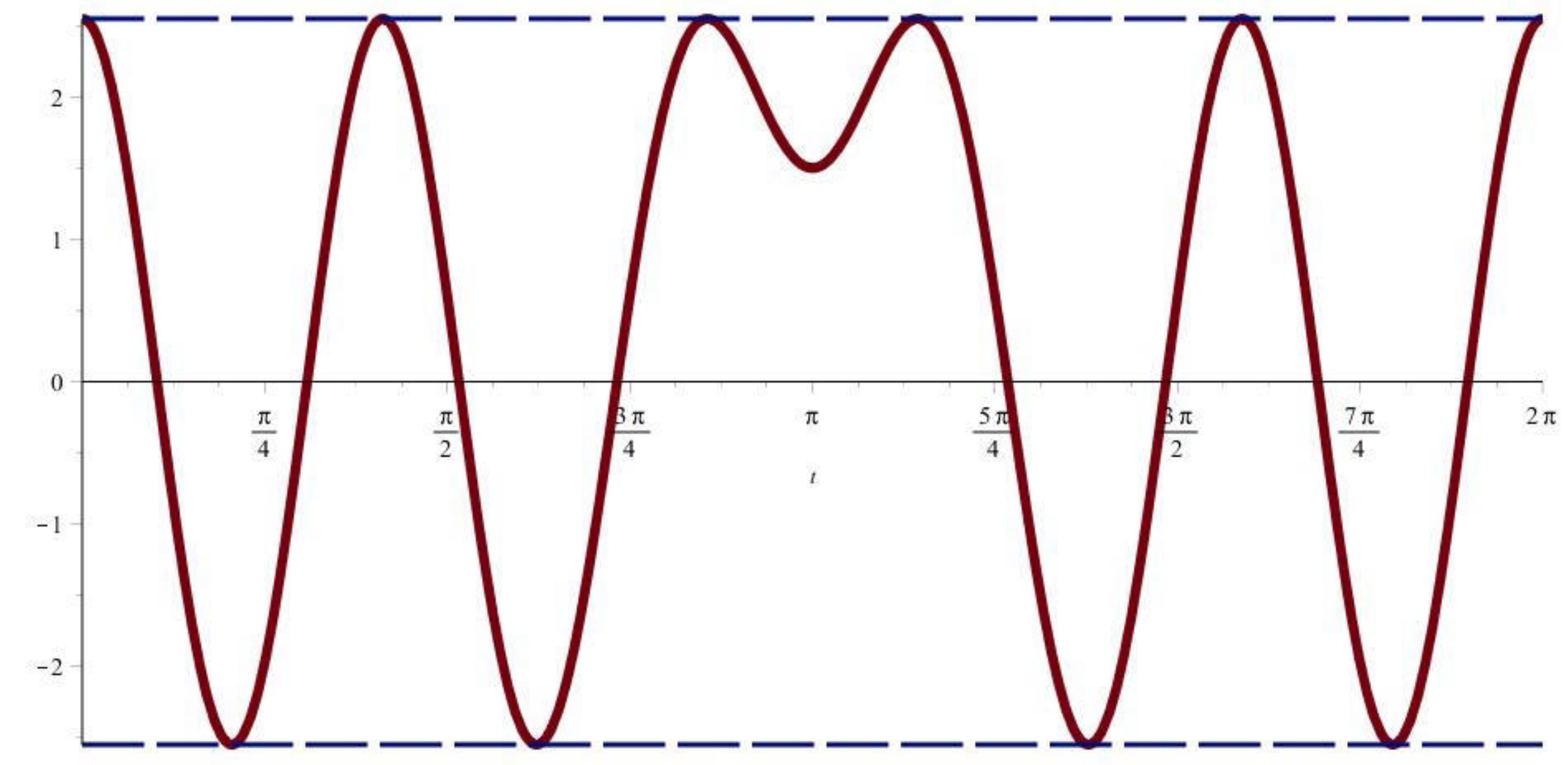}~~~~~~~~
         \includegraphics[width=70mm]{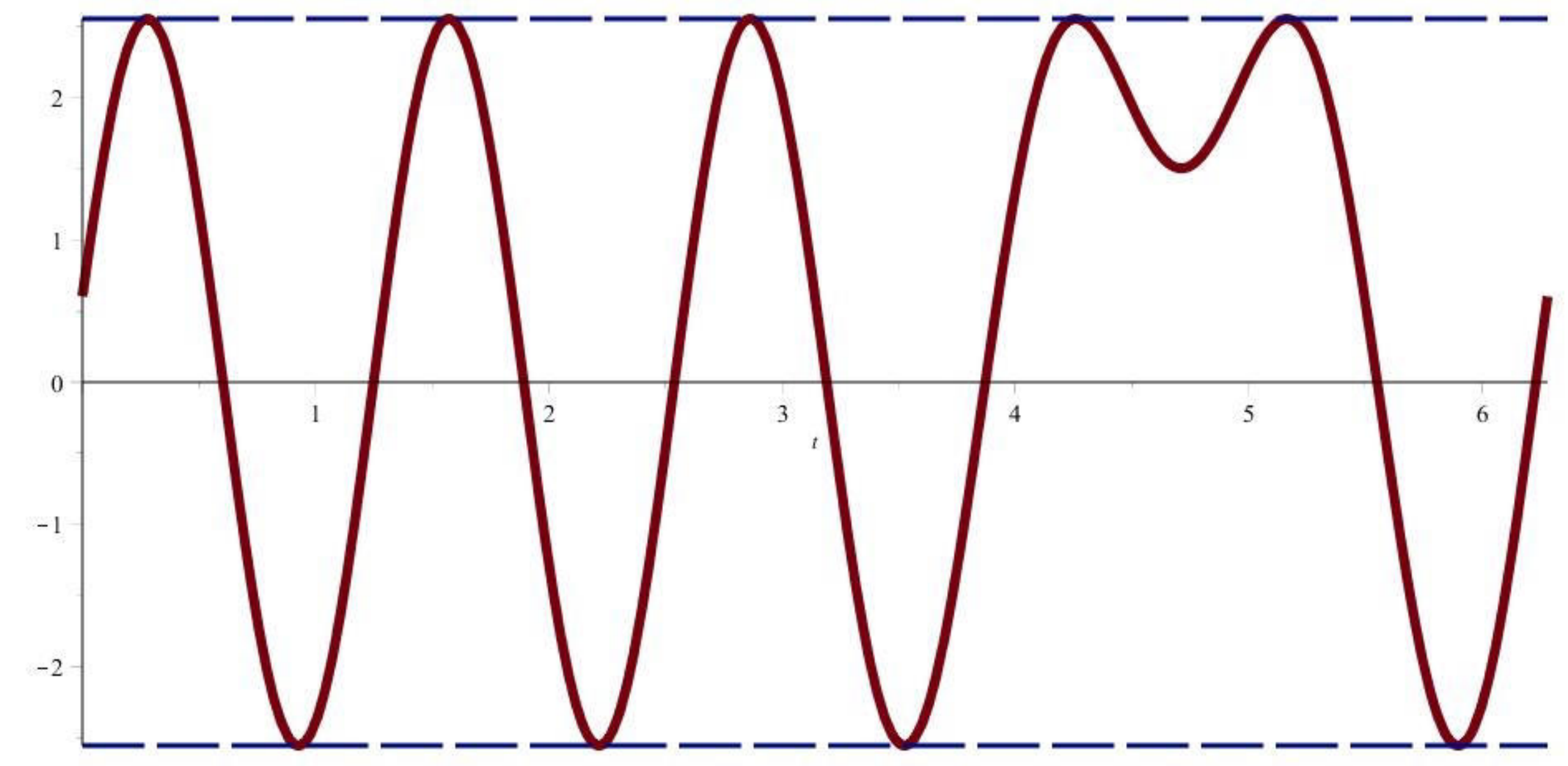}
     \end{center}
       \caption{\it The  function $\psi^*$ of  Theorem \ref{Theorem3.1} for two Fourier regression models of the form \eqref{n1} and \eqref{n2} with $m=5$.
       Left part:  design $\xi_1^*$  of Theorem \ref{Theorem4.1}
       ($b_0=1$, $b_1=0$, $b_2=2$), right part:    design $\widetilde{\xi}^*_1$ of Theorem \ref{Theorem4.2}  ($b_0=1$, $b_1=2$, $b_2=0$).    \label{fig02}}
    \end{figure}

\bex
Consider the  case $m = 5$, $b_1= 0$, $b_2= 2$ and $k_1=4$, $k_2=3$.
 The $T$-optimal discriminating design can be obtained from Theorem \ref{Theorem4.1} and is given by
 $$
 \xi_1^* = \begin{pmatrix}
0 & 0.65 & 1.29 & 1.95 & 2.69 & 3.59 & 4.33 & 4.99 & 5.64\\
0.20 & 0.18 & 0.13 & 0.07 & 0.02 & 0.02 & 0.07 & 0.13 & 0.18
\end{pmatrix}
 $$
 Similarly, if $b_1=2, b_2=0$ the $T$-optimal discriminating design is given by
 $$ \tilde \xi_1^* = \begin{pmatrix}
1.57 & 2.21 & 2.86 & 3.52 & 4.26 & 5.16 & 5.9 & 0.28 & 0.93 \\
0.20 & 0.18 & 0.13 & 0.07 & 0.02 & 0.02 & 0.07 & 0.13 & 0.18
\end{pmatrix}
 $$
 Note that the   design $ \tilde \xi^*_1$ is obtained from   the design $\xi^*_1$  by the transformation $x \to x + \frac {\pi}{2}$. In Figure \ref{fig02} we display the function $\psi^*$ in the equivalence Theorem \ref{Theorem3.1} for both cases.
\eex

\begin{remark}{\rm  \label{remneu}
In the case $k_1=m-2, k_2=m-1$ explicit solutions can be obtained by similar arguments as given in the proof of Theorem \ref{Theorem4.1} and \ref{Theorem4.2}.
If $ m=2d$ is  even and  $b_1=0$
the function $\overline{\eta}$ is given by
 \begin{eqnarray} \label{2.4aa}
\overline{\eta}(x,q,\overline{b}) = q_{0}+\sum_{i=1}^{2d-2}q_{2i-1}\sin(ix)+\sum_{i=1}^{2d-1}q_{2i}\cos(ix) +
  \sin((2d-1)x)+b_{2}\cos(2dx) ~.
\end{eqnarray}
If  $b_2 \geq \frac {1}{2m}\cot (\frac {\pi}{2m})$, the $T$-optimal discriminating design for the Fourier regression models \eqref{1.3} and \eqref{1.4} with $k_1=m-2$ and $k_2=m-1$ is given by the design \eqref{neu1}, where the support points and
weights are defined by
\begin{eqnarray*}
t^{(1)}_i&=&t^{(\xi_1)}_i+\frac{3\pi}{2}\ \mod\ 2\pi ; ~~ i =1,2 , \ldots, 2m-1~,
\\
  \ww^{(1)}_i &=& \ww^{(\xi_1)}_i; ~~ i=1,2 , \ldots, 2m-1~,
\end{eqnarray*}
respectively, and $t^{(\xi_1)}_i$ and $ \ww^{(\xi_1)}_i;$ are the support points of the design $\xi_1^*$  in \eqref{optdes1}.
The extremal polynomial $\psi^*$ in Theorem \ref{Theorem3.1} is given by
\begin{eqnarray*}
\psi^*(x)=\overline{\eta}(x,\theta^{*},\overline{b})=(-1)^{\frac{m}{2}} |b_1|\Big(1+\frac{1}{2m|b_1|}\Big)^{m}T_m\Big(\frac{-\sin(x)+\frac{1}{2m|b_1|}}{1+\frac{1}{2m|b_1|}}\Big),
\end{eqnarray*}
where the fact that $\psi^*$ can be represented in the form \eqref{2.4aa} follows from \eqref{arcos}.
A similar result is available in the case $b_2 <0, |b_2|\geq \frac {1}{2m} \cot (\frac {\pi}{2m})$ and the details are omitted for the sake of brevity.
}

\end{remark}

\section{\bf Some numerical results}\label{section4}
\def\theequation{4\arabic{equation}}
\setcounter{equation}{0}

The results of Section \ref{Section3b} are only correct  if the module of $b_1$ or $b_2$  is larger or equal to some threshold.
Otherwise  $T$-optimal designs have a more complicated structure and  have to be  found numerically [see \cite{detproshihar2015} for some
algorithms]. In this section we provide some more insight in the structure of    $T$-optimal discriminating designs in    cases, where an analytical determination
of the optimal design is not possible.   For this purpose we consider the Fourier regression  models \eqref{n1} and  \eqref{n2},
where $b_0=1$ and  $b_1,\ b_2\neq 0$. Recalling the representation \eqref{2.4a} for
 the function $\overline{\eta}$ in \eqref{1.5}, we see that the support points  and weights of the optimal  $T$-discriminating designs depend on the two parameters $b_1, b_2$ of the extended model. Moreover, the structure of the optimal design changes and depends on the
location  of the point ($b_1,b_2$).
We have calculated  $T$-optimal discriminating designs for the Fourier  regression  models \eqref{n1} and  \eqref{n2}
for $m=2$ and $m=3$.

\begin{figure}[hhh!]
      \begin{center}
        \includegraphics[width=78mm]{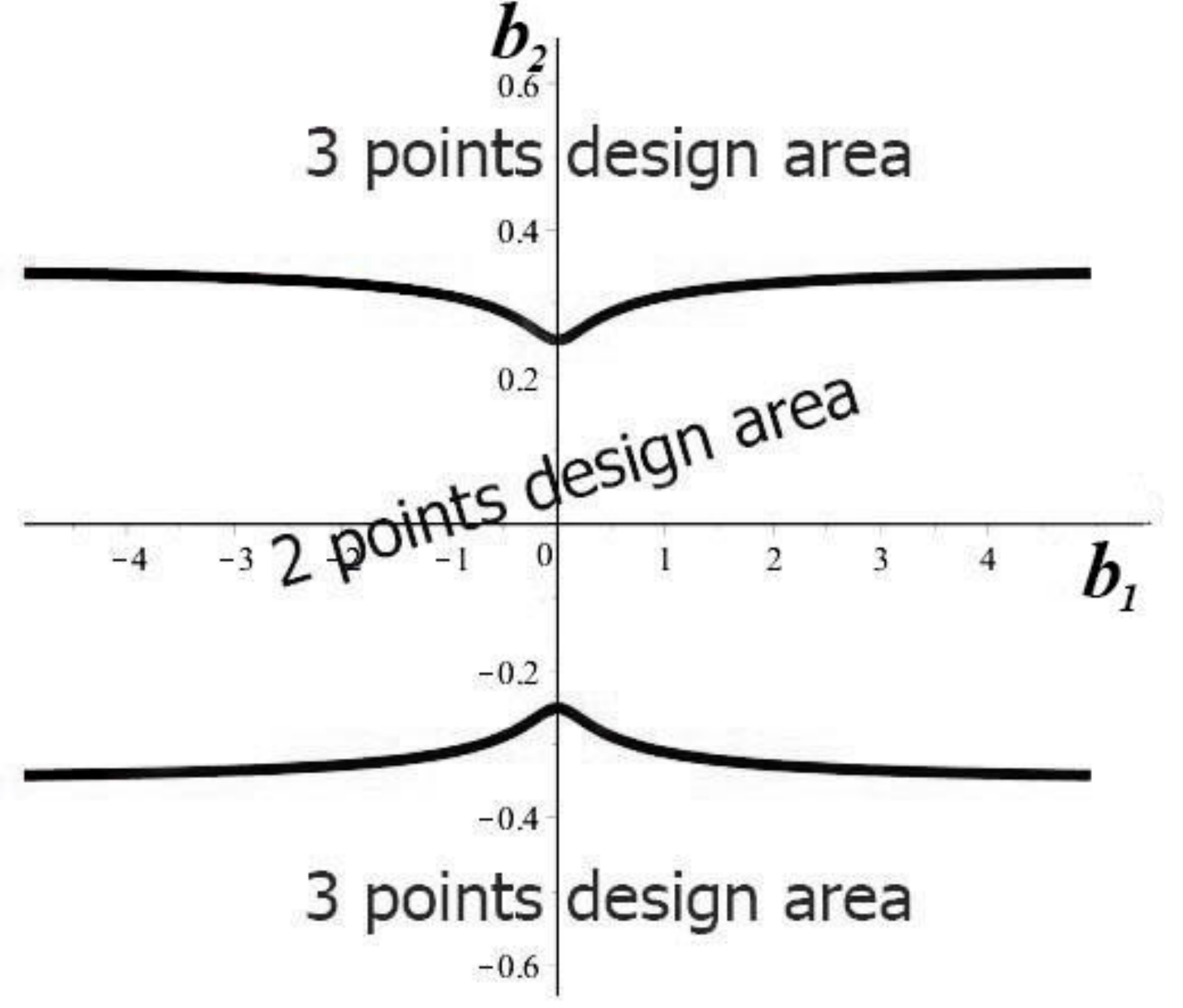} ~~~
        \includegraphics[width=78mm]{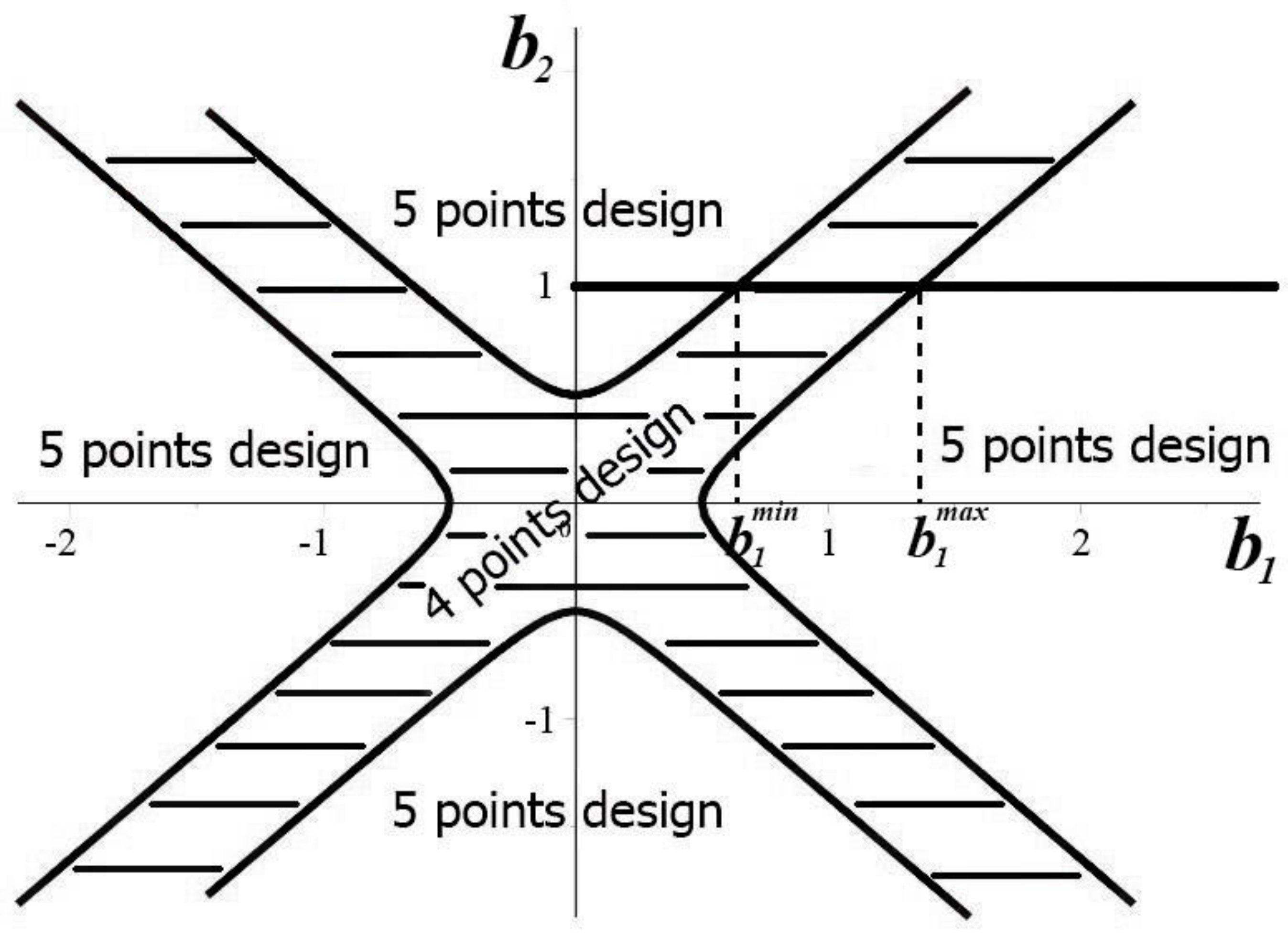}
      \end{center}
       \caption{ \it The  number of support points of the    $T$-optimal design for the Fourier regression models \eqref{n1}
       and  \eqref{n2}. Left part: $m=2$, right part: $m=3$.}\label{fig04}
    \end{figure}

If $m=2$ the  $T$-optimal designs have either $2$ or $3$ support points, and the corresponding areas for the point $(b_1,b_2)$ are depicted in the left part of
 Figure \ref{fig04}. For example, if $b_1=0$ and $|b_2| \geq 0.25$ the locally $T$-optimal discriminating design has $3$ support points (which coincides with the results
 of Theorem \ref{Theorem4.1}), while in the opposite case the optimal design is supported at only two points.  This pattern does not
 change if $b_1 \not =0$, but the threshold is slightly increasing.
  Numerical calculations show that the threshold converges to   $\frac{\sqrt{2}}{4}$ as $b_1 \to \infty$.

\begin{figure}[hhh!]
      \begin{center}
        \hskip-1em \includegraphics[width=75mm,height=60mm]{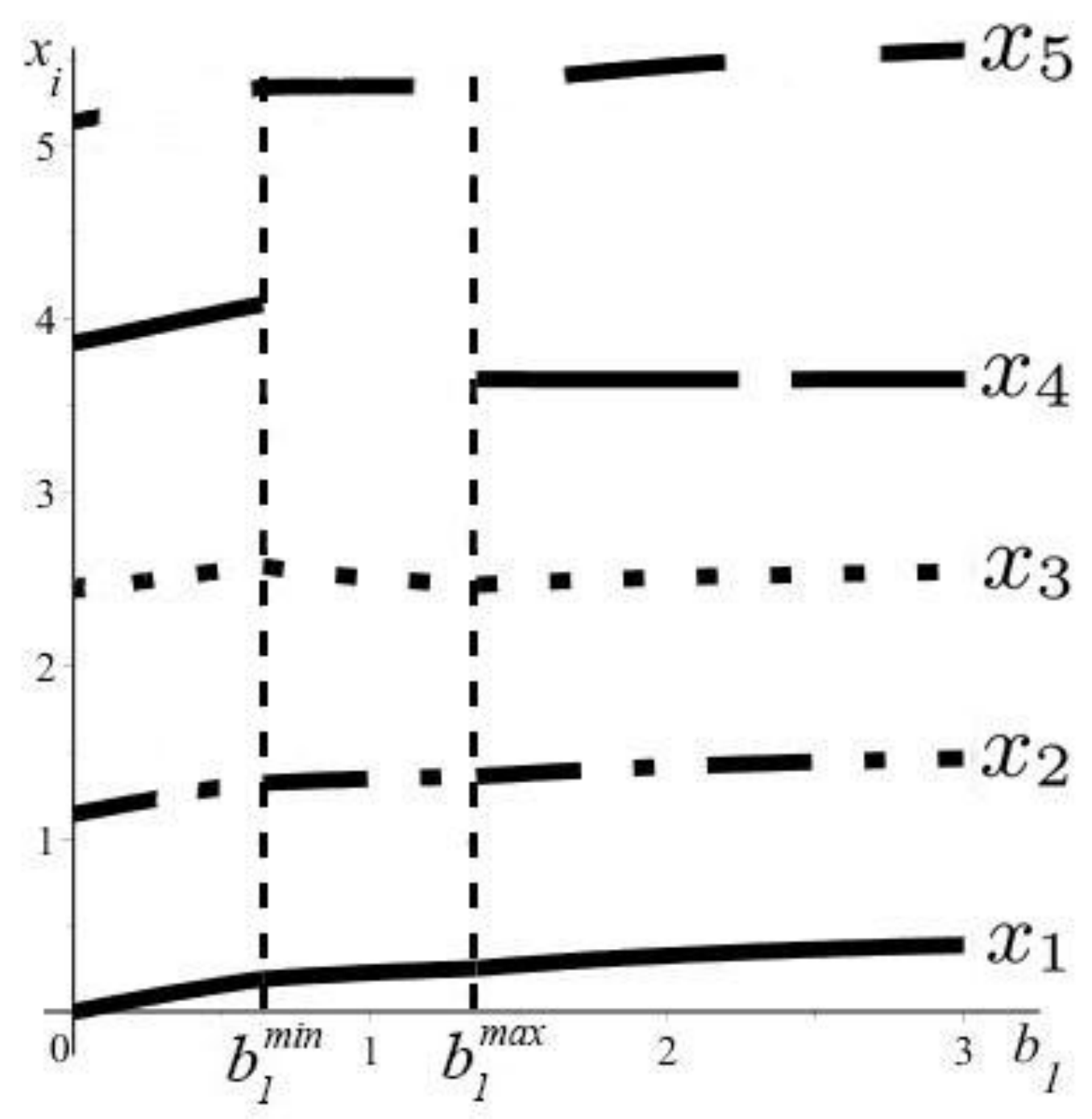} ~~~
         \includegraphics[width=75mm,height=60mm]{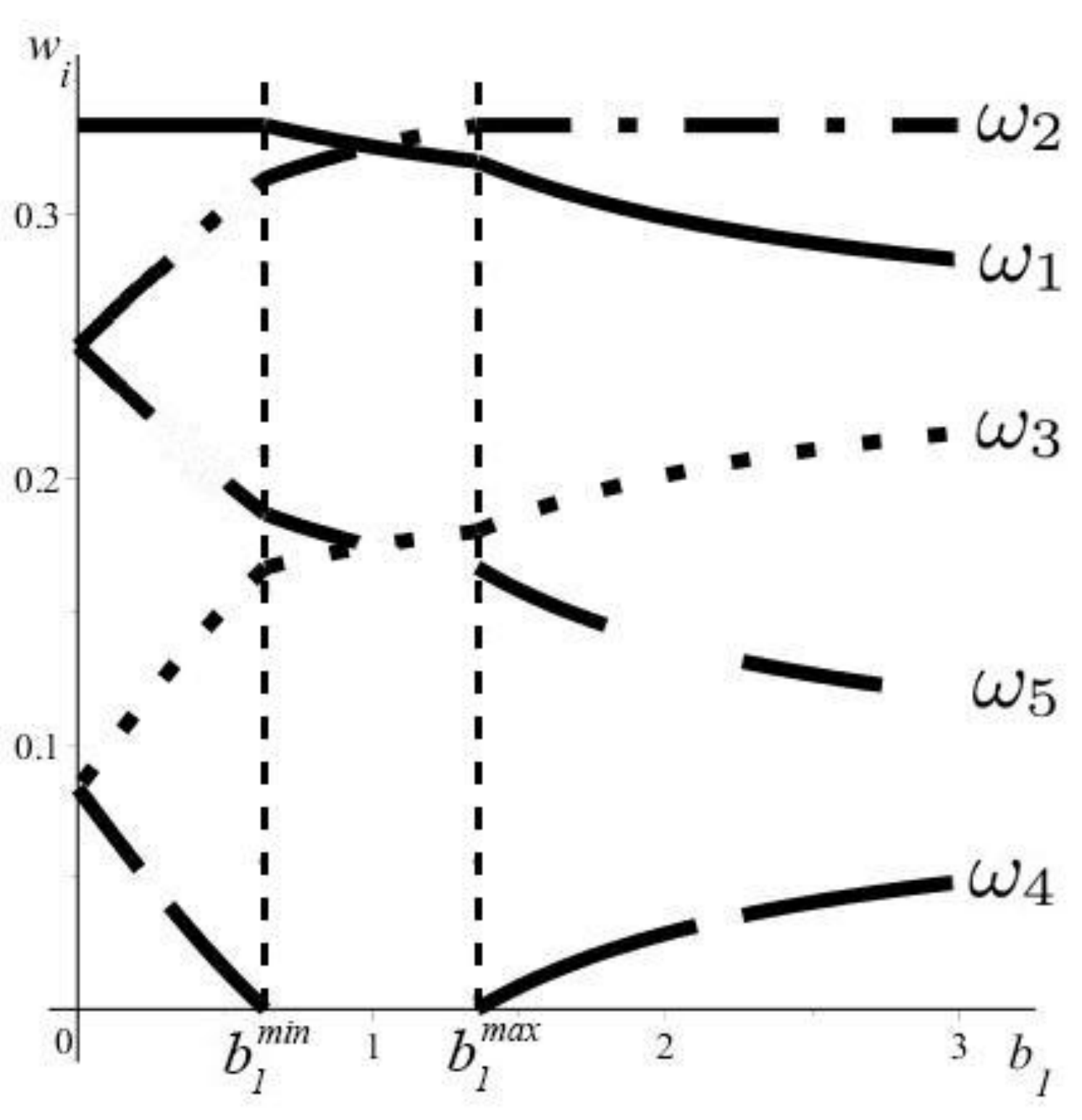}\hskip-2em
      \end{center}
       \caption{ \it The support points and weights of the   $T$-optimal discriminating design for the Fourier regression models
       \eqref{n1} and \eqref{n2}, where $m=3$, $b_0=1$, $b_2=1$,  and  $b_1\in [0,3]$.}\label{fig06}
    \end{figure}

    The right part of Figure \ref{fig04} shows corresponding results for the case $m=3$, and we see that the plane is separated into five
    parts. Four of them correspond to parameter configurations, where the  $T$-optimal discriminating design is supported at $5$ points. Additionally,
     there exists one component, where a $4$-point design is  $T$-optimal for discriminating between the two Fourier regression models.
Consider for example  the situation, where $b_2=1$ and $b_1$ varies in the interval $[0, 3]$. In this case there exist two values, say $b_1^{min}$ and
$ b_1^{max}$, where the line through the point $(0,1)$ in the direction $(1,0)$ intersects the boundary of the fourth region  [see the  right part of Figure \ref{fig04}]. If $b_1 \in  [0, b_1^{min}]$ the $T$-optimal discriminating design
has $5$ support points, while it has only $4$ support points if  $b_1 \in   [b_1^{min}, b_1^{max}]$. Finally, on  the interval $[b_1^{max}, 3]$ the
$T$-optimal discriminating design has again $5$ support points. The support points and corresponding weights of the
$T$-optimal discriminating design are shown in   Figure \ref{fig06} [for the Fourier regression models \eqref{n1} and \eqref{n2}] as a function of the parameter $b_1 \in  [0, 3]$ where $b_2=1$.

\begin{figure}[hhh!]
      \begin{center}
          \hskip0em \includegraphics[width=85mm,height=50mm]{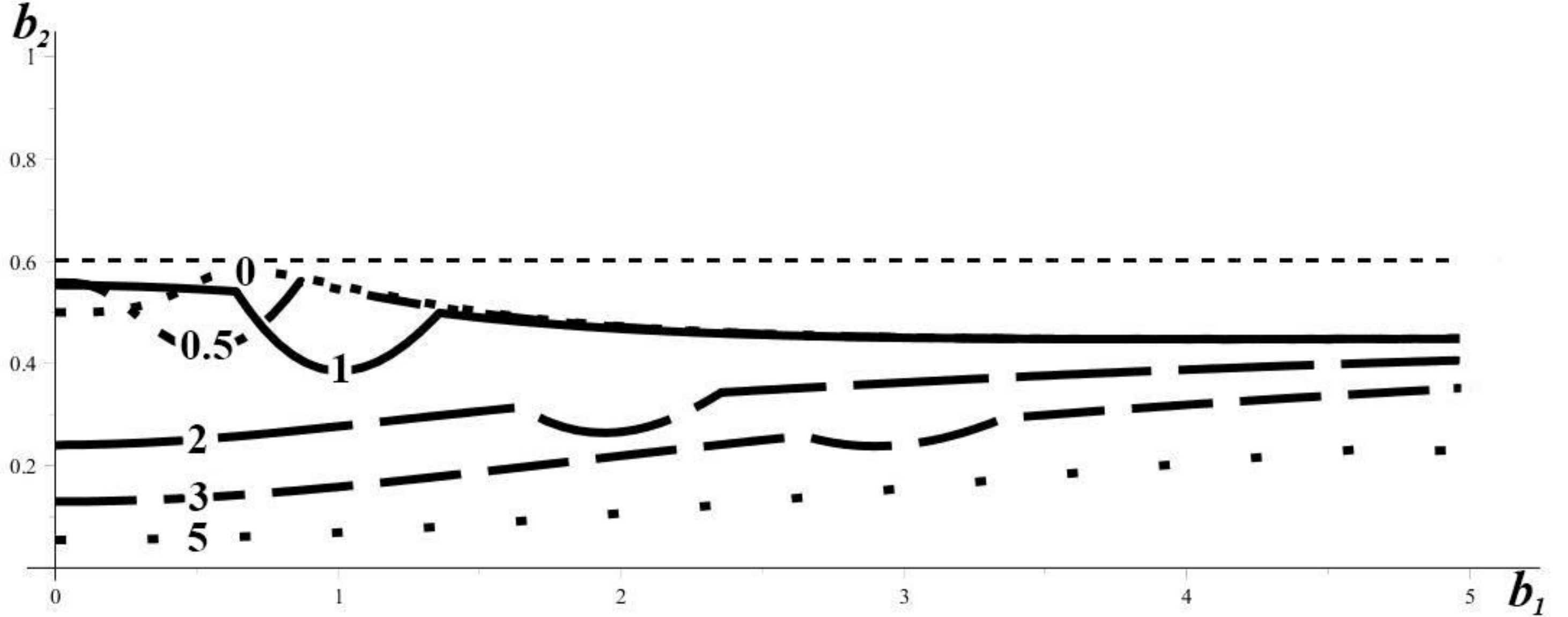} ~~\includegraphics[width=85mm,height=50mm]{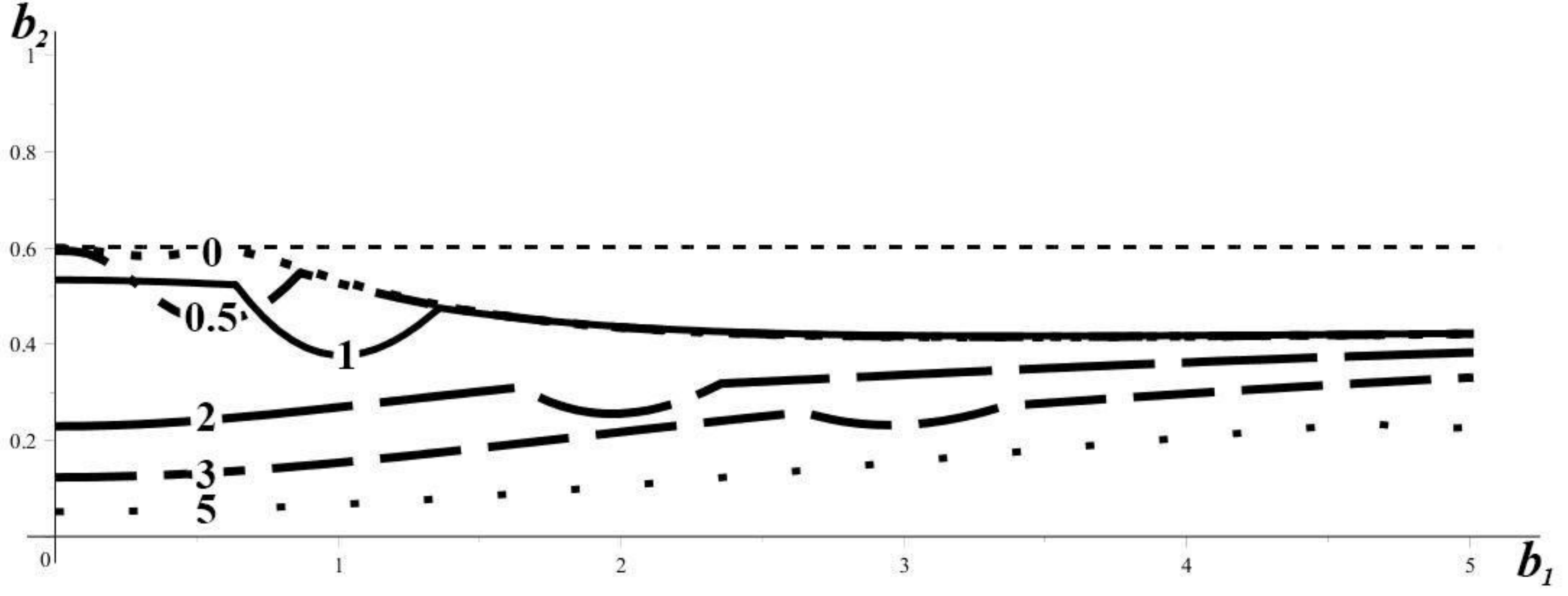}
      \end{center}\vskip-0.75em
       \caption{ \it The $T$-efficiency of the $D$-optimal design (left part) and $D_3$-optimal design (right part)
       for discriminating between the Fourier regression models \eqref{n1} and \eqref{n2}, where  $m=3$, $b_2= 0,0.5,1,2,3,5$,
       $b_1\in [0,5]$.}\label{fig08}
    \end{figure}

We conclude this section investigating the $T$-efficiency
\begin{eqnarray*}
 \mbox{Eff}_T(\xi,b)=\frac{T(\xi,b)}{\max_{\eta}T(\eta,b)}
\end{eqnarray*}
of some commonly used designs in this context. The first design
is the $D$-optimal design for the extended model \eqref{1.4}. The design  can be found in  \cite{pukelsheim2006} and is given by
$$
\xi_D^* =\begin{pmatrix}
0 & \frac{\pi}{4} & \frac{\pi}{2} & \frac{3\pi}{4} & \pi & \frac{5\pi}{4} & \frac{3\pi}{2} & \frac{7\pi}{4} \\
 & & & & & &\\
\frac{1}{8} &  \frac{1}{8} & \frac{1}{8}  &  \frac{1}{8} & \frac{1}{8}  &  \frac{1}{8} & \frac{1}{8} &  \frac{1}{8}
\end{pmatrix}
$$
The second design
is a  discriminating design in the sense of  \cite{stigler1971}.   This design provides a most accurate estimation of the
 three highest coefficients $b_0, b_1$ and $b_2$ in model \eqref{n2} and can be obtained from the results of \cite{laustu1985}. The design
is given by
 \begin{eqnarray*}
  \xi^*_{D_3}
  =\begin{pmatrix}
0 & \frac{\pi}{4} & \frac{\pi}{2} & \frac{3\pi}{4} & \pi & \frac{5\pi}{4} & \frac{3\pi}{2} & \frac{7\pi}{4} \\
 & & & & & &\\
\frac{3}{20} &  \frac{1}{10} & \frac{3}{20}  &  \frac{1}{10} & \frac{3}{20}  &  \frac{1}{10} & \frac{3}{20} &  \frac{1}{10}
\end{pmatrix}
\end{eqnarray*}
and  will be called $D_3$-optimal design throughout this section. The corresponding efficiencies are shown in Figure \ref{fig08}
for various values of $b_2$, where the parameter $b_1$ varies in the interval $[0,5]$.
Both designs have rather similar $T$-efficiencies which are always smaller than $60\%$.
 This similarity can be explained by the fact  that the $D$- and $D_3$-optimal design have the same support and only differ with respect to their weights.
 The efficiencies are decreasing with the parameter $b_2$.
For larger values of $b_2$ the efficiencies of the $D$-and $D_3$-optimal design are very low. For fixed
$b_2$ and larger values of $b_1$ the efficiencies do not change substantially.

\bigskip
\bigskip

{\bf Acknowledgements} The authors would like to thank Martina
Stein, who typed parts of this manuscript with considerable
technical expertise. The work  of V.B. Melas and  P. Shpilev was supported by St. Petersburg State University (project  "Actual problems of design and
analysis for regression models", 6.38.435.2015).
This work has also been supported in part by the
Collaborative Research Center ``Statistical modeling of nonlinear
dynamic processes'' (SFB 823, Teilprojekt C2) of the German Research Foundation
(DFG) and  by a grant from the National Institute Of General Medical Sciences of the National
Institutes of Health under Award Number R01GM107639. The content is solely the responsibility of the authors and does not necessarily
 represent the official views of the National
Institutes of Health.

{  \small
\bibliographystyle{apalike}
 \setlength{\bibsep}{ 2pt}
\bibliography{TrigonomT-optimal17122015}
}

\end{document}